
\input phyzzx.tex
\pubnum={HUPD-9305}
\date={January 1993}

\titlepage
\title{Two-loop effective potential in quantum field theory in curved
space-time.}

\author{S.D. Odintsov}

\address{Tomsk Pedagogical Inst.,\break
 634041 Tomsk, Russia \break
and\break
 Dept. of Physics,\break
 Hiroshima University,\break
 Higashi-Hiroshima 724,\break
JAPAN \footnote{1)}{temporal postal address;}  \footnote{2)}
{email address: odintsov@theo.phys.sci.hiroshima-u.ac.jp}}

\abstract{The method of the calculation of effective potential (in linear
curvature approximation and at any loop) in massless gauge theory in curved
space-time by the direct solution of renormalization group equation is given.
The closed expression for two-loop effective potential is obtained.  Two-loop
effective potential in $\lambda\varphi^4$-theory is written explicitly.  Some
comments about it as well as about two-loop effective potential in standard
model are presented.}
\sequentialequations
 The effective potential [1] plays very important role in the study of vacuum
structure in modern quantum field theory.  The renormalization group(RG)
method turns out to be very useful in calculation and discussion of effective
potential in loops expansion [1].  In particular, using one-loop effective
potential which is reliable only for limited range of background scalar field
and using the fact that the effective potential satisfies the RG equation one
can sum leading (or subleading) logarithms [1]  (see also [2-6] for extension
and applications to different field models, and [5] for a general review.) In
other words, we can extend the range of scalar field.  This is important, for
a example, for a discussion of stability of electroweak vacuum [5,6] (see also
[10]).

{}From another side, RG method can help us to calculate the effective potential
by direct solution of RG equation if the corresponding $\beta$-functions are
known [1].  In particular, the knowledge of the two-loop $\beta$-functions for
a general gauge model [7-9] gives us the way to calculate the two-loop
effective potential [13].  This fact is quite remarkable because the direct
tedious calculation of two-loop effective potential even in scalar
self-interacting
theory in flat space started long ago [11] has been completed quite recently
[12,14].

The effective potential in quantum field theory in curved space-time (for a
general review, see [15]) is quite important for applications in early
Universe. In particular, this potential is required for analysis of GUT
inflation and chaotic inflation (for a general review see [16,17]) where
curvature effects are quite essential. (Actually we need the effective action
 but the calculation of the effective potential can be regarded as the first
step in this direction.)

The RG method for calculation of one-loop effective potential in arbitrary
massless gauge theory in curved space has been developed in ref.[18] (see [19]
for RG improved potential).  In the present letter the method of the
calculation of the effective potential at any loop by direct solution of RG
equation in massless gauge theory in curved space-time is given. (We work in
linear curvature approximation).  This method is simple extension of the
corresponding method in flat space [13].  As an example, we find the two-loop
effective potential in $\lambda\varphi^4$-theory and in standard model.

Let us consider an arbitrary renormalizable massless gauge theory with the
scalars $\varphi$, spinors and gauge fields in curved space-time.  Let
$\tilde{g} \equiv (g,\lambda,h)$ where $g$ is Yang-Mills coupling, $\lambda$
and $h$ are scalar and Yukawa couplings respectively.  The tree effective
potential has the following form:
$$
V^{(0)} \equiv V^{(0)}_\lambda + V^{(0)}_R = a\lambda\varphi^4-b\xi R\varphi^2
\eqno\eq
$$
where $a,b$ are some numerical constants.  In what follows we work in linear
curvature approximation taking into account only such gravitational
corrections.

The RG equation for effective potential is [15,18]
$$
 (\mu {\partial \over {\partial\mu}}+\beta_{\tilde{g}} {\partial \over
{\partial \tilde{g}}}+\delta{\partial \over {\partial \alpha}}+\beta_{\xi}
{\partial \over{\partial\xi}}-\gamma\varphi{\partial \over {\partial\varphi}})
V=0
\eqno\eq
$$
Working in Landau gauge in which $\delta=0$ we can drop $\delta{\partial \over
{\partial\alpha}}$ term from Eq.(2).

Let us suppose that $V_\lambda$ and $V_R$ satisfy the equation (2)
independently. (Then of course, $V=V_\lambda + V_R$ also satisfies it). Then
following the ref.[13] we can write the following n-loop order RG equation for
effective potential:
$$
(\mu{\partial \over {\partial\mu}} V^{(n)}+D_n V^{(0)}+D_{n-1}V^{(1)}+ \cdots
+D_1 V^{(n-1)})=0
\eqno\eq
$$
where $D_n = \beta^{(n)}_{\tilde{g}} {\partial \over {\partial\tilde{g}}}
+\beta^{(n)}_\xi {\partial \over{\partial \xi}}-\gamma^{(n)}\varphi {\partial
\over{\partial\varphi}}$, $V^{(n)}$ is the n-loop correction to effective
potential, $\beta^{(n)}$ is the n-loop correction to the corresponding
$\beta$-function.  Note that in accord with our proposal we have two equations
(3)-one for $V_\lambda$ and one for $V_R$.

Now one can find the effective potential\footnote{1)}{Note that direct
calculation of one-loop effective potential (usually in $\lambda\varphi^4$-
theory) in some definite spaces like $S_4$, $R_3 \times S_1$, etc. has been
given in many works (for review and references see [15]).} by using the
recursion formula (3).  Actually, for $V_\lambda$ it has been done in ref.[13]
up to two loops.  On the same way one can find $V_R$ up to two loops.  The
result can be written in the following form (we will give here two-loop
potential only):

$$
\eqalign{
 V &= V^{(0)}+ V^{(1)} + V^{(2)} \cr
&= a\lambda\varphi^4 + A^{(1)} \varphi^4 ln {\varphi^2 \over {\mu^2}}+{1\over2}
   \{[\beta^{(2)} _\lambda - 4\lambda\gamma^{(2)} ]a
   -2\gamma^{(1)} A^{(1)} \} \times \varphi^4 ln {\varphi^2 \over {\mu^2}} \cr
&+{1\over 4} \{\beta^{(1)}_
   {\tilde g} [{\partial A^{(1)} \over \partial {\tilde g} }]
   -4 \gamma^{(1)} A^{(1)} \} \varphi^4 (ln {\varphi^2 \over {\mu^2}})^2
\cr
&-b\xi R \varphi^2-B^{(1)} R \varphi^2 ln {\varphi^2 \over {\mu^2}}
-{1\over2}(\beta^{(2)}_\xi -2\xi\gamma^{(2)} - 2\gamma^{(1)}{B^{(1)}\over b})
   bR\varphi^2 \ln {\varphi^2 \over {\mu^2}}
\cr
&-{1\over4} \{ \beta^{(1)}_{\tilde g}
  { \partial B^{(1)} \over \partial {\tilde g} }
  +\beta^{(1)}_\xi{  \partial B^{(1)} \over \partial \xi }
  -2\gamma^{(1)}B^{(1)} \} R\varphi^2 (\ln {\varphi^2 \over {\mu^2}})^2 }
\eqno\eq
$$
where $A^{(1)}= {a \over 2}[\beta^{(1)}_\lambda - 4\lambda\gamma^{(1)}],\ \
B^{(1)}={b\over 2}[\beta ^{(1)}_\xi - 2\xi\gamma^{(1)}].$

This is our main result-two-loop effective potential in linear curvature
approximation.  Note that this is the first example of the calculation of the
two-loop effective potential in curved space-time.  Note also that we found
the effective potential using the following normalization conditions:

$$
\eqalign{
&V^{(1)}_\lambda \big\vert_{\mu=\varphi} = 0,\ \ \ \
 V^{(2)}_\lambda \big\vert _{\mu=\varphi} =0, \cr
&V^{(1)}_R \big\vert _{\mu=\varphi} =0,\ \ \ \  V^{(2)}_R
\big\vert_{\mu=\varphi}=0
\cr
}\eqno\eq
$$

These conditions have been used in flat space in [5,13], they are slightly
different from the Coleman-Weinberg normalization conditions [1]. Of course,
non-logarithmic terms may be changed by chosing of other normalization
conditions.  However, leading logarithmic terms are independent of it [12].  In
principle, there is no problem to generalize (4) for higher loops too.

Now we can give some examples.  For $\lambda\varphi^4$-theory with $a={1\over
24}, b={1 \over 2}$ we have the two-loop $\beta$-functions [12,14] in the form:

$$
\eqalign{
&\beta^{(1)}_\lambda+\beta^{(2)}_\lambda={3\lambda^2 \over {(4\pi)^2}}- {17
\over 3}{\lambda^3 \over {(4\pi)^4}},\cr
&\gamma^{(1)}+\gamma^{(2)}=0+{\lambda^2 \over {12(4\pi)^4}} \cr
}\eqno\eq
$$
 The RG function for conformal coupling has the following structure [15]
$$
\beta_\xi = (\xi-{1\over 6})\gamma_m^2+\triangle\beta_\xi
\eqno\eq
$$
where $\gamma_m^2$ is the $\gamma$-function for scalar field mass, $\triangle
\beta_\xi$ does not depend on $\xi$, and appears only at two-loop level (and
higher).  For the theory under consideration we have [15,20]
$$
\beta^{(1)}_\xi + \beta^{(2)}_\xi = {{\lambda(\xi-{1\over 6})} \over
{(4\pi)^2}} - {{5\lambda^2(\xi-{1\over 6})} \over {6(4\pi)^4}} + {\lambda^2
\over {18(4\pi)^4}}
\eqno\eq
$$
Substituting (6), (8) into (4) one gets the two-loop effective potential in
$\lambda\varphi^4$-theory:
$$
\eqalign{
V&={\lambda \over 24}\varphi^4 - {1\over2}\xi R \varphi^2 +
{{\lambda^2 \varphi^4}\over{(16\pi)^2}} \ln{{\varphi^2}\over{\mu^2}}
- {{\lambda(\xi-{1\over6})}\over{(8\pi)^2}}R\varphi^2 \ln{\varphi^2
\over {\mu^2}}
\cr
& -{{\lambda^3\varphi^4} \over {8(4\pi)^4}} \ln {\varphi^2 \over {\mu^2}}+
{{3\lambda^3 \varphi^4} \over {32(4\pi)^4}} (\ln {\varphi^2 \over {\mu^2}})^2
\cr
&
-{{\lambda^2} \over {4(4\pi)^4}}[(\xi-{1\over 6})+ {1\over 36}]R \varphi^2
\ln {\varphi^2 \over {\mu^2}} - {\lambda^2(\xi-{1\over 6}) \over {4(4\pi)^4}}
R \varphi^2 (\ln {\varphi^2 \over {\mu^2}})^2
}\eqno\eq
$$
It is interesting that $R \varphi^2(ln {\varphi^2 \over {\mu^2}})^2$ term
disappears if $\xi={1\over 6}$.

On the same way one can find two-loop effective potential in other theories.
For a example in standard model (we use notations of ref.[14]) $a={1\over 24},
b={1\over 2},$ and two-loop $\beta$-functions for $\lambda$, gauge couplings
$g, g', g_3$ and top quark Yukawa coupling $h$ are given in refs. [9,14].  The
only problem is connected with $\beta_{\xi}$ (more exactly, with
$\triangle\beta_\xi$ as $\gamma_m^2$ is also given in [14]).  The scalar loops
correction to $\triangle\beta^{(2)}_\xi$ ($\triangle\beta^{(1)}_\xi=0$ always)
is given by the expression

$$
\triangle\beta^{(2)}_\xi(scalar)={\lambda^2 \over {9(4\pi)^4}}
\eqno\eq
$$
The gauge and Yukawa couplings contributions to $\triangle\beta^{(2)}_\xi$ are
unknown and should be calculated (this is quite difficult task).  However,
there are some indications [15]  that gauge and Yukawa couplings corrections
happen to break the conformal invariance in third loop, or even higher.  Then
we may conjecture that there are no any additional corrections to
$\triangle\beta^{(2)}_\xi$ excepting (10).  ( Of course, this conjecture should
be checked).  Substituting all these $\beta$-functions to (4) one can get the
two-loop effective potential for standard model in curved space-time.  (We
don't write it explicitly because it is straitforward but the final expression
has extremely long form).

Let us give finally few remarks about two-loop effective potential (19) in
$\lambda\varphi^4$-theory.  One can see that if $\xi R > 0, \lambda > 0$ then
there exists the spontaneous symmetry breaking
$$
\varphi^2 = 6\xi R/\lambda
\eqno\eq
$$
Using (9), from ${\partial V \over {\partial \varphi}}=0$ one can find the
two-loop corrections to vacuum (11) in this case.

Even in the case $\xi = 0$, one can see that for $\varphi^2$ there can exist
the minimum (spontaneous symmetry breaking) defined by trancendental equation
${\partial V \over {\partial \varphi}}=0$.

The two-loop effective potential is useful also for description of
curvature-induced phase transitions [15].  This is known that such transition
can be important for inflationary universe.  In particular, in
asymptotically-free theories one-loop approach is not enough [18] to answer
whether such phase transition is possible.  This question can be answered in
frames of above approach if we would know $\beta^{(2)}_\xi$.

The other interesting extension of this work may be the discussion of massive
theories along the same line.  However, in this case the linear curvature
approach is not enough (actually, we need also all $R^2$-corrections) and whole
story is going to be much more couplicated from technical viewpoint.

 I would like to thank I. Antoniadis and T. Muta for the interest in this work.
 Financial support from JSPS, Japan is greatly appreciated. I am also grateful
to Emi Nakamoto for typing the manuscript.


\REF\Wei{E. Weinberg and S. Coleman, {\it Phys. Rev.} {\bf D7} (1973) 1888.}
\REF\Ein{M.B. Einhorn and D.R.T. Jones, {\it Nucl. Phys.} {\bf B211} (1983)
29.}
\REF\Wes{G.B. West, {\it Phys. Rev.} {\bf D27} (1983) 1402;\nextline
K. Yamagishi, {\it Nucl. Phys.} {\bf B216} (1983) 508.}
\REF\Ein{M.B. Einhorn and D.R.T. Jones, {\it Nucl. Phys.} {\bf B230} (1984)
261.}
\REF\She{M. Sher, {\it Phys. Repts.} {\bf 179} (1989) 274.}
\REF\Dun{M.J. Duncan, R. Phillipe and M. Sher, {\it Phys. Lett.} {\bf B153}
(1985) 165;\nextline
M. Lindner, M. Sher and H.W. Zaglauer, {\it Phys. Lett.} {\bf B228} (1989)
139;\nextline
J. Ellis, A. Linde and M. Sher, {\it Phys. Lett.} {\bf B252} (1990) 203.}
\REF\Jon{D.R.T. Jones, {\it Nucl. Phys.} {\bf B75} (1974) 531; {\it Phys. Rev.}
{\bf D25} (1982) 581.}
\REF\Jac{I. Jack and H. Osborn, {\it J. Phys.}{\bf A16} (1983) 1101; {\it Nucl.
Phys.} {\bf B234} (1983) 331; {\bf B249} (1985) 472 \nextline
M.E. Machacek and M.T. Vaughn, {\it Nucl. Phys.} {\bf B249} (1985) 70.}
\REF\For{C. Ford, I. Jack and D.R.T. Jones, {\it Nucl. Phys.} {\bf B387} (1992)
373.}
\REF\Cab{N. Cabibbo, L. Maiani, G. Parisi and R. Petronzio, {\it Nucl. Phys.}
{\bf B158} (1979) 295.}
\REF\Lee{S.Y. Lee and A.M. Sciccaluga, {\it Nucl. Phys.} {\bf B96} (1975) 435.}
\REF\For{C. Ford and D.R.T. Jones, {\it Phys. Lett.} {\bf B274} (1992) 409.}
\REF\Alh{H. Alhendi, {\it Phys. Rev.} {\bf D37} (1988) 3749.}
\REF\Ford{C.Ford, D.R.T. Jones, P.W. Stephenson and M.B. Einhorn, preprint
 LTH288.}
\REF\Buc{I.L. Buchbinder, S.D. Odintsov, I.L. Shapiro, Effective action in
quantum gravity, IOP Publishing, Bristol and Philadelphia, 1992.}
\REF\Lin{A.D Linde, Particle Physics and Inflationary Cosmology, Harwood
Academic, NY 1990.}
\REF\Kol{E.W. Kolb and M.S. Turner, The early universe, Addison-Wesley, 1990}
\REF\Buc{I.L. Buchbinder, S.D. Odintsov, {\it Class. Quant. Grav.} {\bf 2}
(1985) 721;\nextline
S.D. Odintsov and I.L. Shapiro, {\it Class. Quant. Grav.} {\bf 9} (1992) 873.}
\REF\Eli{E. Elizalde and S.D. Odintsov, preprint HUPD 93-04, 1993.}
\REF\Bun{T. Bunch and L. Parker, {\it Phys. Rev.} {\bf D20} (1979) 2499.}
\refout
\bye